\documentstyle[11pt,epsf,fleqn]{article}
\begin{document}
\title{Electroweak phase transition in a nonminimal supersymmetric model}
\author{S.W. Ham$^{(1)}$,  S.K. Oh$^{(1,2)}$, C.M. Kim$^{(2)}$, E.J. Yoo$^{(2)}$, D. Son$^{(1)}$
\\
\\
{\it $^{\rm (1)}$ Center for High Energy Physics, Kyungpook National University}\\
{\it Daegu 702-701, Korea} \\
{\it $^{\rm (2)}$ Department of Physics, Konkuk University, Seoul 143-701, Korea}
\\
\\
}
\date{}
\maketitle
\begin{abstract}
The Higgs potential of the minimal nonminimal supersymmetric
standard model (MNMSSM) is investigated within the context of
electroweak phase transition. We investigate the allowed parameter
space yielding correct electroweak phase transitoin employing a
high temperature approximation. We devote to phenomenological
consequences for the Higgs sector of the MNMSSM for
electron-positron colliders. It is observed that a future $e^+
e^-$ linear collider with $\sqrt{s} = 1000$ GeV will be able to
test the model with regard to electroweak baryogenesis.
\end{abstract}
\vfil
\eject

\section{INTRODUCTION}

The baryogenesis via electroweak phase transition is one of the
most interesting scenarios for explaining the baryon asymmetry of
the universe, and it can be tested in the future accelerator
experiments [1, 2, 3]. The mechanism for dynamically generating
the baryon asymmetry at the electroweak scale should satisfy the
three Sakharov conditions, namely, the presence of baryon number
violating processes, the violation of both C and CP, and a
departure from thermal equilibrium [4]. If the electroweak phase
transition is responsible for the baryon asymmetry, it should be
first order in order to satisfy the third Sakharov condition.
Furthermore, the strength of the first order phase transition
should be very strong in order to be sufficiently far away from
thermal equilibrium. Otherwise, the baryon asymmetry might be
washed out. Thus, the strongly first order phase transition
requires that the vacuum expectation value (VEV) in the broken
phase should be larger than the critical temperature.

In order to satisfy the requirement that the electroweak phase transition should be
strongly first order, the standard model is found to predict a light Higgs boson,
which is much smaller than the experimental lower bound set by the LEP2 data [5].
On the other hand, in order the standard model to predict the Higgs boson mass larger
than the experimental constraint of LEP2, the electroweak phase transition should be weakly first order.
However, in this case, the standard model might not sufficiently generate the baryon asymmetry.
Therefore, within the context of the baryogenesis via the electroweak phase transition,
it is natural to examine possible extensions of the standard model to avoid this difficulty.

One of the most plausible extensions is to embed the supersymmetry
in the standard model. The simplest extension is called the
minimal supersymmetric standard model (MSSM), which has just two
Higgs doublets in its Higgs sector.
In the MSSM, the strength of the electroweak phase transition might be stronger than that of
the standard model, at least for a specific region of the parameter space [6].
Thus, a strongly first order electroweak phase transition is allowed in this parameter region, where
$\tan\beta$ is small, the right-handed stop quark is light, and one of the scalar Higgs boson is light.
Whereas the mass of the scalar Higgs boson is above the experimental constraint of LEP2,
the mass of the right-handed stop quark is predicted to be smaller
than that of top quark in this parameter region.

Introducing a Higgs singlet into the MSSM, we have the next-to-minimal supersymmetric standard model (NMSSM) [7],
where a term which is cubic in the Higgs singlet is present  in the tree-level Higgs potential
in order to account for the breaking of the Peccei-Quinn symmetry which leads to a massless pseudoscalar Higgs boson.
In the NMSSM, due to the presence of the trilinear soft supersymmetry breaking terms
in the tree-level Higgs potential, the first order electroweak phase transition can be sufficiently strong
in a wide region of the parameter space, where the lightest scalar Higgs boson mass can be as large as 170 GeV [8].
It is much easier to get a strongly first order electroweak phase transition in the NMSSM,
consistent with phenomenological constraints, compared to the MSSM [8, 9, 10].
Also, a U(1)$'$ extension of the MSSM has been investigated in Ref. [11], where a strong enough
first order electroweak phase transition for electroweak baryogenesis is shown to exist.

Recently a extension of the MSSM is constructed in a minimal way [12].
It is called the minimal nonminimal supersymmetric standard model (MNMSSM).
In the MNMSSM, the cubic term of the Higgs singlet is absent from the superpotential of the MNMSSM.
The explicit breaking of the Peccei-Quinn symmetry is accomplished by the presence of
the singlet tadpole term arising from higher order interactions.
The dimensional $\mu$ parameter of the MSSM is generated dynamically by means of the VEV of the neutral Higgs singlet.
A number of investigations on the phenomenology of the MNMSSM have been done,
for example, concentrating on Higgs production [13], spontaneous CP violation [14],
and electroweak baryogenesis [15] in its Higgs sector.

In this article, we consider the possibility of examining the electroweak phase transition
in the MNMSSM at the future $e^+e^-$ linear colliders.
First, we check if the electroweak phase transition in the MNMSSM may be strongly first order.
We calculate the critical temperature where two minima of the potential are degenerate, and express
the Higgs boson masses in terms of the critical temperature.
For the parameter region where the electroweak phase transition is strongly first order,
the production cross section for the neutral Higgs boson is calculated in $e^+ e^-$ collisions,
such as LEP2 and the future linear colliders.

The article is organized as follows.
In the next section, the Higgs sector of the MNMSSM is presented.
In Sec. III, we calculate critical temperatures and the VEVs of the Higgs fields at them.
In Sec. IV, the Higgs boson masses and their productions in $e^+ e^-$ collisions are
obtained at critical temperature.
Discussions and conclusions are given in Sec. V.

\section{THE HIGGS POTENTIAL}

The Higgs sector of the MNMSSM consists of two Higgs doublets $H_i$ $(i = 1, 2)$ and a Higgs singlet $N$,
where the superpotential is given by
\[
    W = h_t Q H_2 t_R^c + h_b Q H_1 b_R^c + \lambda N H_1^T \epsilon H_2 \ ,
\]
where the first two terms are analogous to the Yukawa interaction terms for
the third generation quarks in the MSSM.
The $2\times2$ antisymmetric matrix $\epsilon$ is defined as $\epsilon_{12} = - \epsilon_{21} = 1$.
The third term corresponds to the so called $\mu$-term of the MSSM,
when the Higgs singlet develops the VEV.
Thus, all the coupling coefficient in the above superpotential are dimensionless.
There is no term that is cubic in the Higgs singlet in the MNMSSM.

The tree-level Higgs potential of the MNMSSM may be decomposed into $D$ terms, $F$ terms,
and soft supersymmetry breaking terms as
\begin{equation}
        V_0 = V_D + V_F + V_{\rm S} \ ,
\end{equation}
where
\begin{eqnarray}
    V_D & = & {g_2^2\over 8} (H_1^{\dag} \vec\sigma H_1
        + H_2^{\dag} \vec\sigma H_2)^2 + {g_1^2 \over 8} (|H_2|^2 - |H_1|^2)^2  \ , \cr
    V_F & = & |\lambda|^2 [(|H_1|^2 + |H_2|^2) |N|^2 + |H_1^T \epsilon H_2|^2] \ , \cr
    V_{\rm S} & = & m_{H_1}^2 |H_1|^2 + m_{H_2}^2 |H_2|^2 + m_N^2 |N|^2
            - (\lambda A_\lambda H_1^T \epsilon H_2 N + {\rm H.c.})\ .
\end{eqnarray}
Here, $\vec\sigma$ are the Pauli matrices, and $m_{H_1}^2$, $m_{H_2}^2$, and $m_N^2$ are
the soft SUSY breaking masses, and $A_{\lambda}$ is the trilinear soft SUSY breaking parameter with mass dimension.

The above tree-level Higgs potential has a Peccei-Quinn symmetry, which leads us to an unwanted axion
after spontaneous symmetry breakdown, since the determinant of the pseudoscalar Higgs boson mass matrix is zero.
In order to get rid of this axion in the MNMSSM, we introduce a tadpole term to the above potential
due to higher-order interactions for the Higgs singlet [11].
The tadpole term may be given linear in the Higgs singlet as
\begin{equation}
    V_{\rm tadpole}  =  - \xi^3 N - \xi^3 N^* \ ,
\end{equation}
where $\xi$ is the tadpole coefficient.
This tadpole coefficient is intrinsically a free parameter, but, for a phenomenological point of view,
we assume that it is of the order of the soft SUSY breaking mass ($\le$ 1 TeV).
Thus, by means of the Higgs singlet tadpole term in the MNMSSM, the Peccei-Quinn symmetry is explicitly broken down.

The one-loop corrections to the tree-level Higgs potential are obtained by considering
the effective potential method [16].
We assume that loops involving only the quarks and scalar quarks for the third generation are sufficient.
If the scalar quark masses are degenerate, the one-loop effective potential
due to the quark and scalar quark for the third generation is given as [17]
\begin{equation}
    V_1 = {3 h_t^4 |H_2|^4 \over 16 \pi^2} \left \{ {3 \over 2}
        + \log \left ( {{\tilde m}^2 + h_t^2 |H_2|^2 \over h_t^2 |H_2|^2} \right ) \right \}
        + {3 h_b^4 |H_1|^4 \over 16 \pi^2} \left \{ {3 \over 2}
        + \log \left ({{\tilde m}^2 + h_b^2 |H_1|^2 \over h_b^2 |H_1|^2 } \right ) \right \} \ ,
\end{equation}
where the soft SUSY breaking mass ${\tilde m}$ satisfies ${\tilde m}^2 = ({\rm 1000 \ GeV})^2 \gg m_q^2$ $(q = t, b)$.
In general, $\lambda$, $A_{\lambda}$, and $\xi$ may be complex numbers.
However, as we allow no CP violation in the Higgs sector, they are taken real.
Further, for simplicity, we take them positive, since negative values yield phenomenologically the same results.

Now, in order to take into account the electroweak phase transition,
we introduce the temperature-dependent contributions to the effective potential [18].
It is well known that in the standard model the high temperature approximation is
consistent with the exact calculation of the integrals to better than 5 \% for $m_f/T < 1.6$
in the case of the fermion and for $m_b/T < 2.2$ in the case of the boson [19].
Thus, we assume that the temperature for the electroweak phase transition in the MNMSSM is high enough
as compared with the masses of the relevant particles.
The temperature-dependent potential comes from the thermal effects due to the Higgs bosons,
the gauge bosons, the quarks for the third generation, the lighter chargino, and three the light neutralinos.
According to the analysis of Ref. [9], the temperature-dependent potential
in the high temperature approximation is given by
\begin{eqnarray}
    V_T & = &
    {T^2 \over 12} \{ 2 (m_{H_1}^2 + m_{H_2}^2) + m_N^2 + c^2 (|H_1|^2
    + |H_2|^2) + 3 (h_t^2 |H_2|^2 + h_b^2 |H_1|^2) \cr
    & &\mbox{} + 6 \lambda^2 |N|^2 \}  \ ,
\end{eqnarray}
where $T$ is the temperature of the potential and $c^2 = g_1^2 + 3 g_2^2 + 3 \lambda^2 $.
Therefore, the full effective potential at the one-loop level for the electroweak phase transition in the MNMSSM is
\begin{equation}
    V = V_0 + V_{\rm tadpole} + V_1 + V_T \ .
\end{equation}
which is now temperature-dependent.

\section{CRITICAL TEMPERATURE}

Now, let us evaluate the critical temperature and the VEVs of the Higgs fields at the critical temperature
in order to examine the order and strength of the phase transition at the electroweak scale.
We denote the temperature-dependent VEVs of the Higgs fields as
$v_1 = \langle (v_1(T), 0) \rangle = \langle (H_1^0(T), H_1^- (T)) \rangle = \langle H_1 (T) \rangle$,
$v_2 = \langle (0, v_2(T)) \rangle = \langle (H_2^+(T), H_2^0 (T)) \rangle = \langle H_2 (T) \rangle$,
and $x=x(T)= \langle N(T) \rangle$.
Making a ${\rm SU}(2)_L \times {\rm U}(1)_Y$ gauge transformation
one can choose $<H_2^+> = 0$ and $v_2 > 0$.
The condition for a local minimum with $<H_1^-> = 0$ is equivalent to the requirement
that the charged Higgs boson have positive mass.
The temperature-dependent vacuum is defined as the minimum of the effective potential $V$ as
\begin{eqnarray}
    \langle V \rangle
    & = &  {g_1^2 + g_2^2 \over 8} (v_1^2 - v_2^2)^2 + \lambda^2 (v_1^2 + v_2^2) x^2 + \lambda^2 v_1^2 v_2^2
    + m_{H_1}^2 v_1^2         + m_{H_2}^2 v_2^2 + m_N^2 x^2 \cr
    & &\mbox{} - 2 \lambda A_{\lambda} v_1 v_2 x - 2 \xi^3 x + {3 h_t^4 v_2^4 \over 16 \pi^2} \left \{ {3 \over 2} +            \log \left ({{\tilde m}^2 + h_t^2 v_2^2 \over h_t^2 v_2^2} \right ) \right \} \cr
    & &\mbox{} + {3 h_b^4 v_1^4 \over 16 \pi^2} \left \{ {3 \over 2} + \log \left ({{\tilde m}^2 + h_b^2 v_1^2 \over
        h_b^2 v_1^2} \right ) \right \} + {T^2 \over 12} \{ 2 (m_{H_1}^2 + m_{H_2}^2) + m_N^2 \cr
    & &\mbox{} + (v_1^2 + v_2^2) c + 3 (h_t^2 v_2^2 + h_b^2 v_1^2) + 6 \lambda^2 x^2 \} \ .
\end{eqnarray}
Here, the soft SUSY breaking masses at the zero-temperature are given by
\begin{eqnarray}
    m_{H_1}^2
    & = &\mbox{} - {m_Z^2 \over 2} \cos 2 \beta - \lambda^2 (x(0)^2 + v(0)^2 \sin^2 \beta) + \lambda A_{\lambda}
        \tan \beta x(0) \cr
    & &\mbox{}- {3 h_b^2 m_b^2 \over 16 \pi^2} \left \{ 2 + 2 \log \left ( {{\tilde m}^2 + m_b^2 \over m_b^2}\right )
        + {m_b^2 \over {\tilde m}^2 + m_b^2} \right \} \ , \cr
    m_{H_2}^2
    & = & {m_Z^2 \over 2} \cos 2 \beta - \lambda^2 (x(0)^2 + v(0)^2 \cos^2 \beta) + \lambda A_{\lambda}
        \cot \beta x(0) \cr
    & &\mbox{} - {3 h_t^2 m_t^2 \over 16 \pi^2} \left \{ 2 + 2 \log \left ( {{\tilde m}^2 + m_t^2 \over m_t^2}\right )
        + {m_t^2 \over {\tilde m}^2 + m_t^2} \right \} \ , \cr
    m_N^2
    & = &\mbox{} - \lambda^2 v(0)^2 + {\lambda \over 2 x(0)} v(0)^2 A_{\lambda} \sin 2 \beta + {\xi^3 \over x(0)} \ ,
\end{eqnarray}
where $\tan\beta = v_1(0)/v_2(0)$, $v^2(0) = v_1^2(0) + v_2^2(0) $, and $v(0) = v(T=0) = 175$ GeV.
The temperature-dependent vacuum is obtained by solving three minimum equations
\begin{eqnarray}
    0 & = & 2 m_{H_1}^2 v_1 - 2 \lambda A_{\lambda} v_2 x + 2 \lambda^2 v_1 x^2 + 2 \lambda^2 v_1 v_2^2
        +{g_1^2 + g_2^2 \over 2} (v_1^2 - v_2^2) v_1 + {T^2 \over 6} (c^2 + 3 h_b^2) v_1 \cr
    & &\mbox{}
        + {3 h_b^4 v_1^3 \over 8 \pi^2} \left \{ 2 + 2 \log \left ({{\tilde m}^2 + h_b^2 v_1^2 \over h_b^2 v_1^2} \right )      + {h_b^2 v_1^2 \over {\tilde m}^2 + h_b^2 v_1^2} \right \} \ , \cr
    0 & = & 2 m_{H_2}^2 v_2 - 2 \lambda A_{\lambda} v_1 x + 2 \lambda^2 v_2 x^2 + 2 \lambda^2 v_1^2 v_2
        + {g_1^2 + g_2^2 \over 2} (v_2^2 - v_1^2) v_2 + {T^2 \over 6} (c^2 + 3 h_t^2) v_2 \cr
    & &\mbox{}
        + {3 h_t^4 v_2^3 \over 8 \pi^2} \left \{ 2 + 2 \log \left ({{\tilde m}^2 + h_t^2 v_2^2 \over h_t^2 v_2^2} \right )
        + {h_t^2 v_2^2 \over {\tilde m}^2 + h_t^2 v_2^2} \right \} \ , \cr
    0 & = & 2 m_N^2 x - 2 \lambda A_{\lambda} v_1 v_2 + 2 \lambda^2 (v_1^2 + v_2^2) x - 2 \xi^3 + T^2 \lambda^2 x \ ,
\end{eqnarray}
for the three VEVs, namely, $v_1$, $v_2$, $x$, and then substituting them into the expression for $\langle V \rangle$.
The above minimum equations are nonlinear and thus can not be solved exactly by analytical method.
Eventually, in $\langle V \rangle$, we are left with six free parameters as $\tan \beta$, $A_{\lambda}$,
$\lambda$, $x(0)$, $\xi$, and the temperature $T$.

In the NMSSM, the critical temperature is defined as the largest value among the temperatures
which give the second derivative of the Higgs potential zero at the origin of space of fields [8, 9].
In our work, we would like to use a more accurate method to determine the critical temperature
than the method that determines it from the saddle point of the potential.
Our method is to determine the critical temperature as the temperature at which two vacua are degenerate.

Let us employ two independent methods in solving the above nonlinear equations for the three VEVs.
First, we solve them numerically.
From the third equation, one can express $x$ at finite temperature in terms of the other two VEVs as
\begin{equation}
    x = {\lambda A_{\lambda} v_1 v_2 + \xi^3 \over m_N^2 + \lambda^2 (v_1^2 + v_2^2) + \lambda^2 T^2/2} \ .
\end{equation}
Substituting this expression into the remaining two equations, and rewrite them as
\begin{eqnarray}
    && 0 = f_1 (v_1, v_2, T)  \cr
    && 0 = f_2 (v_1, v_2, T)
\end{eqnarray}
where the remaining other free parameters are omitted.
Solving these two equations, one can obtain $v_1$ and $v_2$ for given $T$, and then get $x$.
In practice, we plot $f_1$ and $f_2$ in the ($v_1$, $v_2$)-plane, for a given temperature,
by varying both $v_1$ and $v_2$, with certain values for the remaining free parameters.
We determine the set of points in the ($v_1$, $v_2$)-plane where both of $f_1$ and $f_2$ are zero.
In this way, we determine $v_1$ and $v_2$ for given $T$, and then get $x$.

In Fig. 1a, we plot in the ($v_1$, $v_2$)-plane the contours of $f_1 =0$ and $f_2 = 0$,
for $\tan \beta = 2$, $A_{\lambda} = 504.5$ GeV, $\lambda = 0.3$, $x(0) = 25$ GeV, $\xi = 50$ GeV, and $T = 100$ GeV.
The two contours intersect at five points in the ($v_1$, $v_2$)-plane.
The values of $v_1$ and $v_2$ for these points (and the value of $x$ therefrom) are the solutions of
the above minimum equations.
In order to examine the nature of these points, we plot $\langle V \rangle$ in Fig. 1b
for the same set of parameter values.
One can see in Fig. 1b a number of equipotential contours in the ($v_1$, $v_2$)-plane,
and the five stationary points in the plane where the first derivative of $\langle V \rangle$ is zero.
These points are exactly the five points where both $f_1=0$ and $f_2=0$.
Among them, three points correspond to minima of  the potential,
where its second derivative of the potential is positive, and the other two points correspond to saddle points,
where the sign of the second derivative of the potential is indeterminate.

We compare the values of the three minima of the potential to determine the true global minimum,
where the vacuum is defined.
Among the three minima, the potential at $(v_1, v_2, x) = (2.9, 83.1, 2.0)$ GeV
and $(426.3, 385.0, 234.3)$ GeV have the same value and lower than the one at the remaining point.
Thus, we define two degenerate global minima, that is, the vacua, separated by a potential barrier.
The electroweak phase transition may take place from one vacuum to the other through quantum tunnelling,
and it is first order.
The temperature which we set as 100 GeV is therefore the critical temperature $T_c$.

Let us denote the two points in the ($v_1$, $v_2$, $x$)-space where the two degenerate vacua are
defined as point A and point B. Thus, the coordinates of point A is $(v_{1A}, v_{2A}, x_A) = (2.9, 83.1, 2.0)$ GeV
and those of point B  $(v_{1B}, v_{2B}, x_B) = (426.3, 385.0, 234.3)$ GeV in the ($v_1$, $v_2$, $x$)-space.
We measure the distance between the two points in the ($v_1$, $v_2$, $x$)-space at the critical temperature as
\begin{equation}
    d(T_c) = \sqrt{\{v_{1A} (T_c)-v_{1B} (T_c)\}^2 + \{v_{2A} (T_c)-v_{2B} (T_c)\}^2 + \{x_A (T_c)-x_B (T_c)\}^2} \ .
\end{equation}
As is well known, the requirement for a strongly first order electroweak phase transition is $d(T_c) > T_c$,
In the present case, we have $T_c = 100$ GeV and $d(T_c) = 569.4$ GeV, which yields the ratio $d(T_c)/T_c = 5.6$.
The electroweak phase transition in the present case is therefore safely a strongly first order one.

We repeat the above procedure for other sets of parameter values.
For simplicity, we do not explore the whole parameter space, but vary $\tan \beta$ and $A_{\lambda}$
while fixing the values of the other parameters, namely, $\lambda = 0.3$, $x(0) = 25$ GeV, $\xi = 50$ GeV,
and $T_c = 100$ GeV.
Figs 2a, 3a, 4a, and 5a show the contours of $f_1 = 0$ (solid curve) and $f_2 = 0$ (dashed curve),
and Figs 2b, 3b, 4b, and 5b show the contours of $\langle V \rangle$, for different values of
$\tan \beta$ and $A_{\lambda}$.
The numerical results for these figures, including that of Fig. 1, are summarized by Table I.
As one can see in Table I, the electroweak phase transitions for all the sets of parameters
we consider are strongly first order ones.

Now, as a second and independent way to investigate the electroweak phase transition,
let us employ Newton's method for solving the nonlinear equations, Eq. (9).
The Newton's method starts with a Jacobian matrix with respect to $v_1$, $v_2$, and $x$, which is defined as
\begin{equation}
    J_{ij} (v_1, v_2, x)  = \left \lgroup \matrix
    {{\displaystyle {\partial^2 \langle V \rangle  \over \partial v_1 \partial v_1}},
    {\displaystyle {\partial^2 \langle V \rangle \over \partial v_1 \partial v_2}},
    {\displaystyle {\partial^2 \langle V \rangle \over \partial v_1 \partial x}} \cr
    {\displaystyle {\partial^2 \langle V \rangle \over \partial v_2 \partial v_1}},
    {\displaystyle {\partial^2 \langle V \rangle \over \partial v_2 \partial v_2}},
    {\displaystyle {\partial^2 \langle V \rangle \over \partial v_2 \partial x}} \cr
    {\displaystyle {\partial^2 \langle V \rangle \over \partial x \partial v_1}},
    {\displaystyle {\partial^2 \langle V \rangle \over \partial x \partial v_2}},
    {\displaystyle {\partial^2 \langle V \rangle \over \partial x \partial x}}}
\right \rgroup  \ .
\end{equation}
The elements of the Jacobian matrix are given explicitly as
\begin{eqnarray}
    J_{11} & = & 2 m_{H_1}^2 + 2 \lambda^2 x^2 + 2 \lambda^2 v_2^2 + {g_1^2 + g_2^2 \over 2} (3 v_1^2 - v_2^2)
        + {T^2 \over 6} ( c^2 + 3 h_b^2) \cr
        & &\mbox{} + {3 h_b^4 v_1^2 \over 8 \pi^2}
    \left \{ 2 + 6 \log \left ({{\tilde m}^2 + h_b^2 v_1^2 \over h_b^2 v_1^2} \right )
        + {9 h_b^2 v_1^2 \over {\tilde m}^2 + h_b^2 v_1^2}
        - {2 h_b^4 v_1^4 \over ({\tilde m}^2 + h_b^2 v_1^2)^2} \right \} \ , \cr
    J_{22} & = & 2 m_{H_2}^2 + 2 \lambda^2 x^2 + 2 \lambda^2 v_1^2 + {g_1^2 + g_2^2 \over 2} (3 v_2^2 - v_1^2)
        + {T^2 \over 6} ( c^2 + 3 h_t^2) \cr
        & &\mbox{} + {3 h_t^4 v_2^2 \over 8 \pi^2}
    \left \{ 2 + 6 \log \left ({{\tilde m}^2 + h_t^2 v_2^2 \over h_t^2 v_2^2} \right )
        + {9 h_t^2 v_2^2 \over {\tilde m}^2 + h_t^2 v_2^2}
        - {2 h_t^4 v_2^4 \over ({\tilde m}^2 + h_t^2 v_2^2)^2} \right \} \ , \cr
    J_{33} & = & 2 m_N^2 + 2 \lambda^2 (v_1^2 + v_2^2) + \lambda^2 T^2 \ , \cr
    J_{12} & = &\mbox{} - 2 \lambda^2 A_{\lambda} x + 4 \lambda^2 v_1 v_2 - (g_1^2 + g_2^2) v_1 v_2 \ , \cr
    J_{13} & = &\mbox{} - 2 \lambda^2 A_{\lambda} v_2 + 4 \lambda^2 v_1 x \ , \cr
    J_{23} & = &\mbox{} - 2 \lambda^2 A_{\lambda} v_1 + 4 \lambda^2 v_2 x \ .
\end{eqnarray}
The Newton's method provides us with the stationary points, where the first derivative of the potential vanishes.
The stationary points are then classified as minima, maxima, or saddle points,
according to whether the second derivative of the potential is positive, negative, or indeterminate.
The minima are chosen by imposing on stationary points the two conditions
$J_{ii} > 0$ $(i = 1, 2, 3)$ and $\det (J_{ij}) > 0$.
Among the minima, we select two points, say point A and point B,
where $\langle V(v_{1A}, v_{2A}, x_A) \rangle = \langle V(v_{1B}, v_{2B}, x_B) \rangle$
is the true global minimum of the potential, at a given temperature, in order to define the degenerate vacua.
We check and confirm that the results obtained by the Newton's method coincide
with the numerical results of Table I, for the sets of parameter values we set.

\section{HIGGS BOSON MASS AND PRODUCTION}

In this section, we investigate the possibility of examining the parameter space
where the electroweak phase transition in the MNMSSM may take place for the Higgs sector at electroweak baryogenesis.
We calculate the neutral Higgs boson masses and their production cross sections
in $e^+ e^-$ collisions at the critical temperature.

In the pseudoscalar Higgs sector, there is a neutral Goldstone boson related to $Z$ boson.
At the critical temperature, the elements of the $2 \times 2$ pseudoscalar mass matrix $M_P(T_c)$ are given by
\begin{eqnarray}
    M_{P_{11}} (T_c) & = & {2 \lambda A_{\lambda} x \over \sin 2 \beta}
        + {c^2 \over 6} T_c^2 + {h_t^2 \over 2} \cos^2 \beta T_c^2 + {h_b^2 \over 2} \sin^2 \beta T_c^2 \ , \cr
    M_{P_{22}} (T_c) & = & {\lambda v^2 A_{\lambda} \over 2x} \sin 2 \beta +{\xi^3 \over x} + \lambda^2 T_c^2 \ , \cr
    M_{P_{12}} (T_c) & = & \lambda v A_{\lambda} \ .
\end{eqnarray}
From the pseudoscalar mass matrix, the masses of the pseudoscalar Higgs bosons are obtained as
\begin{equation}
    m_{P_1, P_2}^2 (T_c) = {1 \over 2} \left [ {\rm Tr} {M_P(T_c)} \mp \sqrt{{{\rm Tr} M_P(T_c)}^2
    - 4 {\rm det} {M_P(T_c)}  } \right ]     \ .
\end{equation}
Note that, at zero temperature, the determinant of the pseudoscalar Higgs boson mass matrix
is obtained as $\det \{ M_P (0) \} = 2 \lambda A_{\lambda} \xi^3/\sin 2 \beta$.
If $\xi \neq 0$, the presence of the singlet tadpole term breaks explicitly the Peccei-Quinn symmetry.
On the other hand, if $\xi = 0$, the singlet tadpole term disappears and there
is a massless pseudoscalar Higgs boson at the tree level.
The strength of the Peccei-Quinn symmetry breaking depend on the size of $\xi$.

For the scalar Higgs sector, the elements of the $3 \times 3$ scalar mass matrix $M_S(T_c)$
are given at the critical temperature by
\begin{eqnarray}
    M_{S_{11}} (T_c) & = & m_Z^2 \cos^2 \beta + \lambda A_{\lambda} x \tan \beta + {c^2 \over 6} T_c^2
        + {h_b^2 \over 2} T_c^2 + {3 h_b^2 m_b^2 \over 4 \pi^2}
        \log \left ({{\tilde m}^2 + m_b^2 \over m_b^2} \right ) \cr
    & &\mbox{} + {3 \over 8 \pi^2} \left \{ {4 h_b^2 m_b^4 \over {\tilde m}^2 + m_b^2 }
        - {h_b^2 m_b^6 \over ({\tilde m}^2 + m_b^2)^2} \right \} \ , \cr
    M_{S_{22}} (T_c) & = & m_Z^2 \sin^2 \beta + \lambda A_{\lambda} x \cot \beta + {c^2 \over 6} T_c^2
        + {h_t^2 \over 2} T_c^2  + {3 h_t^2 m_t^2 \over 4 \pi^2}
        \log \left ({{\tilde m}^2 + m_t^2 \over m_t^2} \right ) \cr
    & &\mbox{} + {3 \over 8 \pi^2} \left \{ {4 h_t^2 m_t^4 \over {\tilde m}^2 + m_t^2 }
        - {h_t^2 m_t^6 \over ({\tilde m}^2 + m_t^2)^2} \right \} \ , \cr
    M_{S_{33}} (T_c) & = & {\lambda \over 2 x} v^2 A_{\lambda} \sin 2 \beta + {\xi^3 \over x} + \lambda^2 T_c^2 \ , \cr
    M_{S_{12}} (T_c) & = & (\lambda^2 v^2 - {m_Z^2 \over 2} ) \sin 2 \beta - \lambda A_{\lambda} x  \ , \cr
    M_{S_{13}} (T_c) & = & 2 \lambda^2 v \cos \beta x - \lambda v A_{\lambda} \sin \beta \ , \cr
    M_{S_{23}} (T_c) & = & 2 \lambda^2 v \sin \beta x - \lambda v A_{\lambda} \cos \beta \ .
\end{eqnarray}
From the scalar mass matrix, the masses of the scalar Higgs bosons are obtained as
\begin{eqnarray}
    m_{S_j} (T_c)^2
    & = & {1 \over 3} {\rm Tr} \{M_S(T_c)\} + 2 \sqrt{W}    \cos \left \{{\Theta + 2 j \pi \over 3} \right \}
    \  \  \  (j = 1, \ 2, \ 3)  \ ,
\end{eqnarray}
where
\begin{eqnarray}
    \Theta & = & \cos^{-1} \left ({U \over \sqrt{W^3}} \right)  \ , \cr
    W & = & - {1 \over 18} [{\rm Tr} \{M_S(T_c)\} ]^2 + {1 \over 6} {\rm Tr} \{M_S (T_c) M_S(T_c) \} \ , \cr
    U & = & - {5 \over 108} [{\rm Tr} \{M_S(T_c)\} ]^3
        + {1 \over 12} {\rm Tr} \{M_S(T_c)\} {\rm Tr} \{M_S(T_c) M_S(T_c)\}
        + {1 \over 2} \det \{M_S(T_c)\} \ .
\nonumber
\end{eqnarray}

In $e^+ e^-$ collisions, the four main production channels for the neutral Higgs bosons
in the MNMSSM can be considered as follows [20]:
\begin{eqnarray}
    && \mbox{Higgsstrahlung} : \, e^+e^- \rightarrow Z S_i  \ , \cr
    && \mbox{pair production} : \, e^+ e^- \rightarrow S_i P_j \ , \cr
    && \mbox{$WW$ fusion} : \, e^+e^- \rightarrow {\bar \nu}_e \nu_e S_i \ , \cr
    && \mbox{$ZZ$ fusion} :  \, e^+e^- \rightarrow e^+ e^- S_i \ , \nonumber
\end{eqnarray}
where $S_i$ $(i = 1, 2, 3)$ are three scalar Higgs bosons and  $P_j$ $(j = 1, 2)$ are two pseudoscalar Higgs bosons.
At the center of mass energy of LEP2, the dominant production channels
for the neutral Higgs boson among the four channels are the Higgsstrahlung and pair production process.
As the center of mass energy in $e^+ e^-$ collision increases,
$WW$ and $ZZ$ fusion processes becomes important for the production of the neutral Higgs boson.
We note that the Higgs boson mass as well as the relevant coupling
for the Higgs production depends on the critical temperature.
That is, in $e^+ e^-$ collisions the production cross section
for the neutral Higgs boson is evaluated at the critical temperature.

We first calculate the masses of the five neutral Higgs bosons at the critical temperature.
The results are listed in Table II, for the five sets parameters in Table I.
We then calculate the cross sections for the production of neutral Higgs bosons
via two dominant production channels in $e^+ e^-$ collisions
with $\sqrt{s}$ = 209 GeV (the center of mass energy of LEP2),
and list the results in Table III, for the five sets of parameters in Table I.
We find that the 4th set of parameters and the 5th set of parameters yield
production cross sections well below 100 fb.
If we assume that the discovery limit of 100 fb at LEP2,
the LEP2 data could not put any constraint on the MNMSSM with
either the 4th set of parameters or with the 5th set of parameters, whereas the 1st,
the 2nd, and the 3rd set of parameters yield phenomenologically inconsistent results with the LEP2 data.

We repeat the above calculation for $\sqrt{s}$ = 500 GeV (LC500) and 1000 GeV (LC1000).
Here, we calculate 15 different cross sections for production of the neutral Higgs bosons
via all four channels listed above.
The results are listed respectively in Table IV and Table V.
The numbers for the first three sets of parameters are just for comparison.
We are interested in the numbers for the 4th and the 5th set of parameters.
Assuming that the discovery limits as 20 fb for LC500,
one can see in Table IV that, at LC500, the analysis of neutral Higgs production might determine
whether the 4th set of parameters are acceptable experimentally.
However, the neutral Higgs bosons might not sufficiently be produced for the 5th set of parameters,
and thus the test of the 5th set of parameters cannot be done at LC500.
The MNMSSM with the 5th set of parameters may eventually be examined at LC1000,
where some of the neutral Higgs boson production cross sections are above the discovery limit of LC1000,
which we assume to be 5 fb.

Now, we explore not only a number of points but a wide region
in the parameter space of the MNMSSM whether a strongly first order electroweak phase transition is possible.
The ranges for the parameters of the MNMSSM are set as $2 \le \tan \beta \le 40$,
$0 < \lambda \le 0.7$, $0 < A_{\lambda} \le 1500$ GeV, and $0 < x(0), \xi \le 300$ GeV.
The critical temperature is determined by a temperature at which two degenerate global vacua exist.
We search randomly in the region and discovered 5000 points in the parameter space
where the electroweak phase transition is first order by using the Newton's method.
Among them, we obtain 1955 points where the electroweak phase transition satisfies $d(T_c)/T_c > 1$,
that is, the transition is strongly first order.
For each of these 1955 points, we calculate the masses of the neutral Higgs bosons
and the 15 cross sections for their productions at LC1000.
The results are shown in Figs. 6 and 7.
In Fig. 6, we plot the distributions of those 1955 points in the  $(m_{S_1}, m_{P_1})$-plane.
There is a tendency that the lower bound on $m_{P_1}$ depends on $m_{S_1}$.
In Fig. 7, the largest one among 15 production cross sections for the neutral Higgs bosons
is plotted against $m_{S_1}$ for each of the 1955 points.
For reasonable values of $m_{S_1}$, we see that at least one of the 15 production cross sections
is larger than 20 fb.
Thus, we expect that there is an ample opportunity to discover at LC1000
at least one of the five neutral Higgs bosons in the MNMSSM within the context of
the strongly first order electroweak phase transition.

\section{CONCLUSIONS}

We have investigated the possibility of testing at the future $e^+e^-$ colliders
the MNMSSM which accommodates the strongly first order electroweak phase transition.
Both the MNMSSM and the NMSSM have two Higgs doublets and a singlet.
But they differ in that the Peccei-Quinn symmetry is broken in the MNMSSM
by the singlet tadpole term whereas it is broken in the NMSSM by a cubic term in the Higgs singlet.
In the MNMSSM, we have found that 5 sets of parameter values for
which the electroweak phase transition may take place in strongly first order.
Radiative corrections due to the one-loop contribution from the third generation
are taken into account, where scalar top quark masses are assumed to be degenerate.
Finite temperature effects are considered by the one-loop finite-temperature correction
in the high temperature approximation.
Some sets of parameter values for the MNMSSM yield large enough cross sections for production
of some neutral Higgs bosons in $e^+e^-$ collisions at $\sqrt{s}$ = 500 and 1000 GeV.
By exploring not only the 5 sets of parameter values but a wide region
in the parameter space of the MNMSSM, we have discovered that 1955 points in the region provide
strongly first order transitions.
For these points, we have evaluated the masses of the neutral Higgs bosons,
and calculated the cross sections for their productions in the future $e^+e^-$ collisions
at $\sqrt{s}$ = 1000 GeV via four dominant channels.
We have found that reasonable masses are obtained for $S_1$, and the cross sections are
considerably larger than the discovery limit of 5 fb at LC1000.
Therefore, it is possible to test the MNMSSM at LC1000,
within the context of the strongly first order electroweak phase transition.

\vskip 0.3 in

\noindent
{\large {\bf Acknowledgments}}
\vskip 0.2 in
\noindent
This work was supported by Korea Research Foundation Grant (2001-050-D00005).

\vskip 0.2 in


\vfil\eject

{\bf Figure Captions}

\vskip 0.3 in
\noindent
Fig. 1a : Contours of $f_1 = 0$ (solid curve) and  $f_2 = 0$ (dashed curve) in the $(v_1, v_2)$-plane
for $\tan \beta = 2$ and $A_{\lambda} = 504.5$ GeV. Other parameters are fixed as $\lambda = 0.3$,
$x(0) = 25$ GeV, and $\xi = 50$ GeV. The critical temperature is set as 100 GeV.
Note that the two contours intersect at five points in the $(v_1, v_2)$-plane.

\vskip 0.3 in
\noindent
Fig. 1b : Equipotential contours of $\langle V \rangle $ with the same parameter values as Fig. 1a.
There are five extrema; two of them are saddle points and the other three are minima.
Two of the three minima have the same value of $\langle V \rangle $,
thus define global minima or degenerate vacua.
The coordinates in unit of GeV of the degenerate vacua at point A and point B in the $(v_1, v_2)$-plane
the value of $x$ therefrom are respectively (2.9, 83.1, 2.0) and (426.3, 385.0, 234.3).

\vskip 0.3 in
\noindent
Fig. 2a : The same as Fig. 1a, but with different values of $\tan\beta$ and $A_{\lambda}$: $\tan \beta = 3$
and $A_{\lambda} = 433$ GeV.

\vskip 0.3 in
\noindent
Fig. 2b : The same as Fig. 1b, but with different values of $\tan\beta$ and $A_{\lambda}$: $\tan \beta = 3$
and $A_{\lambda} = 433$ GeV.
The coordinates in unit of GeV of the degenerate vacua in the $(v_1, v_2)$-plane and the value of $x$
therefrom are respectively (4.2, 104.8, 3.5) and (375.7, 364.9, 238.7).

\vskip 0.3 in
\noindent
Fig. 3a : The same as Fig. 1a, but with different values of $\tan\beta$ and $A_{\lambda}$:
$\tan \beta = 10$ and $A_{\lambda} = 401$ GeV.

\vskip 0.3 in
\noindent
Fig. 3b : The same as Fig. 1b, but with different values of $\tan\beta$ and $A_{\lambda}$:
$\tan \beta = 10$ and $A_{\lambda} = 401$ GeV.
The coordinates in unit of GeV of the degenerate vacua in the $(v_1, v_2)$-plane and
the value of $x$ therefrom are respectively (5.0, 125.8, 10.7) and (307.7, 364.7, 360.7).

\vskip 0.3 in
\noindent
Fig. 4a : The same as Fig. 1a, but with different values of $\tan\beta$ and $A_{\lambda}$:
$\tan \beta = 20$ and $A_{\lambda} = 499$ GeV.

\vskip 0.3 in
\noindent
Fig. 4b : The same as Fig. 1b, but with different values of $\tan\beta$ and $A_{\lambda}$:
$\tan \beta = 20$ and $A_{\lambda} = 499$ GeV.
The coordinates in unit of GeV of the degenerate vacua in the $(v_1, v_2)$-plane and
the value of $x$ therefrom are respectively (3.6, 128.2, 14.6) and (298.0, 409.0, 526.6).

\vskip 0.3 in
\noindent
Fig. 5a : The same as Fig. 1a, but with different values of $\tan\beta$ and $A_{\lambda}$:
$\tan \beta = 40$ and $A_{\lambda} = 716.5$ GeV.

\vskip 0.3 in
\noindent
Fig. 5b : The same as Fig. 1b, but with different values of $\tan\beta$ and $A_{\lambda}$:
$\tan \beta = 40$ and $A_{\lambda} = 716.5$ GeV.
The coordinates in unit of GeV of the degenerate vacua in the $(v_1, v_2)$-plane and
the value of $x$ therefrom are respectively  (2.1, 129.0, 17.2) and (281.5, 472.7, 786.7).

\vskip 0.3 in
\noindent
Fig. 6 : Plot of the lightest scalar and pseudoscalar neutral Higgs boson masses for
the 1955 points in the parameter space of MNMSSM where the electroweak phase transition
is strongly first order. The points are selected in $2 \le \tan \beta \le 40$, $0 < \lambda \le 0.7$,
$0 < A_{\lambda} \le 1500$ GeV, and $0 < x(0), \xi \le 300$ GeV.

\vskip 0.3 in
\noindent
Fig. 7 : Plot against the mass of the lightest scalar neutral Higgs boson of
the largest of the 15 cross sections for production of any neutral Higgs bosons
via any channel for the 1955 points in Fig. 6, in $e^+e^-$ collisions with $\sqrt{s} = 1000$ GeV.

\vfil\eject

{\bf Table Caption}
\begin{table}[ht]
\caption{The parameter sets of the MNMSSM where strongly first order phase transition occurs.
Other parameters are fixed as $\lambda = 0.3$, $x(0) = 25$ GeV, and $\xi = 50$ GeV.
The critical temperature is set as 100 GeV.}
\begin{center}
\begin{tabular}{c|c|c|c|c|c}
\hline
\hline
parameters & set 1 & set 2 & set 3 & set 4 & set 5  \\
\hline
\hline
$\tan\beta$ & 2 & 3 & 10 & 20 & 40  \\
\hline
$A_\lambda$ (GeV) & 504.5 & 433.0 & 401.0 & 499.0 & 716.5 \\
\hline
corres. figures & Figs. 1 & Figs. 2 & Figs. 3 & Figs. 4 & Figs. 5 \\
\hline
$v_{1A}$ (GeV), $v_{1B}$ (GeV) & 2.9, 426.3 & 4.2, 375.7 & 5.0, 307.7 & 3.6, 298.0 & 2.1, 281.5 \\
\hline
$v_{2A}$ (GeV), $v_{2B}$ (GeV) & 83.1, 385.0 & 104.8, 364.9 & 125.8, 364.7 & 128.2, 409.0 & 129.0, 472.7 \\
\hline
$x_A$ (GeV), $x_B$ (GeV) & 2.0, 234.3 & 3.5, 238.7 & 10.7, 360.7 & 14.6, 526.6 & 17.2, 786.7 \\
\hline
$d(T_c)$ (GeV) & 569.4 & 510.8 & 520.6 & 654.0 & 887.8 \\
\hline
$d(T_c)/T_c$ & 5.6 & 5.1 & 5.2 & 6.5 & 8.8 \\
\hline
\hline
\end{tabular}
\end{center}
\end{table}
\begin{table}[ht]
\caption{The masses of the neutral Higgs bosons in the MNMSSM
with the five sets of parameter values listed in Table I.}
\begin{center}
\begin{tabular}{c|c|c|c|c|c}
\hline
\hline
mass (GeV) & set 1 & set 2 & set 3 & set 4 & set 5 \\
\hline
\hline
$m_{S_1}$ & 46.8 & 49.7 & 67.6 & 74.3 & 76.6 \\
\hline
$m_{S_2}$ & 170.6 & 165.3 & 158.2 & 156.7 & 156.3  \\
\hline
$m_{S_3}$ & 297.8 & 251.6 & 220.8 & 297.0 & 478.1 \\
\hline
$m_{P_1}$ & 65.0 & 61.6 & 70.3 & 75.2 & 77.0 \\
\hline
$m_{P_2}$ & 298.8 & 252.9 & 221.3 & 297.0 & 478.0 \\
\hline
\hline
\end{tabular}
\end{center}
\end{table}
\begin{table}[ht]
\caption{Cross sections for production of the neutral Higgs bosons
in $e^+ e^-$ collision with $\sqrt{s}$ = 209 GeV
via the Higgsstrahlung and the pair production channel with the five sets
of parameter values listed in Table I.}
\begin{center}
\begin{tabular}{c|c|c|c|c|c}
\hline
\hline
cross section (fb) & set 1 & set 2 & set 3 & set 4 & set 5  \\
\hline
\hline
$\sigma (Z S_1)$ & 718.2 & 674.4 & 264.2 & 79.8 & 20.0 \\
\hline
$\sigma (S_1 P_1)$ & 161.9 & 193.5 & 34.0 & 2.9 & 0.1 \\
\hline
\hline
\end{tabular}
\end{center}
\end{table}
\begin{table}[ht]
\caption{Cross sections for production of all neutral Higgs bosons
in $e^+ e^-$ collision with $\sqrt{s}$ = 500 GeV via
the Higgsstrahlung, the pair production, the $WW$ fusion, and the $ZZ$ fusion channel,
with the five sets of parameter values listed in Table I.}
\begin{center}
\begin{tabular}{c|c|c|c|c|c}
\hline
\hline
cross section (fb) & set 1 & set 2 & set 3 & set 4 & set 5  \\
\hline
\hline
$\sigma (Z S_1)$ & 53.5 & 51.0 & 22.7 & 7.3 & 1.8  \\
\hline
$\sigma (Z S_2)$ & 5.6 & 4.0 & 0.03 & 0.01  & 0.01  \\
\hline
$\sigma (Z S_3)$ & 1.7 & 4.6 & 24.8 & 19.4 & 0.0  \\
\hline
$\sigma (S_1 P_1)$ & 30.8 & 36.4 & 8.7 & 0.9 & 0.06  \\
\hline
$\sigma (S_2 P_1)$ & 25.0 & 15.1 & 0.3 & 0.005 & 0.0001  \\
\hline
$\sigma (S_3 P_1)$ & 0.5 & 2.7& 9.0 & 1.8 & 0.0  \\
\hline
$\sigma (S_1 P_2)$ & 0.9 & 3.1 & 8.4 & 1.7 & 0.0  \\
\hline
$\sigma (S_2 P_2)$ & 0.1 & 0.7 & 0.2 & 0.004 & 0.0  \\
\hline
$\sigma (S_3 P_2)$ & 0.0 & 0.0 & 3.6 & 0.0 & 0.0  \\
\hline
$\sigma ({\bar \nu}_e \nu_e S_1)$ & 115.6 & 108.4 & 43.3 & 13.4 & 3.4  \\
\hline
$\sigma ({\bar \nu}_e \nu_e S_2)$ & 5.4 & 4.0 & 0.03 & 0.02 & 0.01  \\
\hline
$\sigma ({\bar \nu}_e \nu_e S_3)$ & 0.6 & 2.5 & 17.1 & 7.7 & 0.002  \\
\hline
$\sigma(e^+ e^- S_1)$ & 12.3 & 11.6 & 4.6 & 1.4 & 0.3 \\
\hline
$\sigma(e^+ e^- S_2)$ & 0.5 & 0.4 & 0.003 & 0.002 & 0.001 \\
\hline
$\sigma(e^+ e^- S_3)$ & 0.07 & 0.2 & 1.8 & 0.8 & 0.0002 \\
\hline
\hline
\end{tabular}
\end{center}
\end{table}
\begin{table}[hb]
\caption{The same as Table IV, but with a higher center of mass energy, $\sqrt{s}$ = 1000 GeV.}
\begin{center}
\begin{tabular}{c|c|c|c|c|c}
\hline
\hline
cross section (fb) & set 1 & set 2 & set 3 & set 4 & set 5  \\
\hline
\hline
$\sigma (Z S_1)$ & 10.7 & 10.3 & 4.6 & 1.5 & 0.3  \\
\hline
$\sigma (Z S_2)$ & 1.4 & 1.0 & 0.008 & 0.004  & 0.003  \\
\hline
$\sigma (Z S_3)$ & 0.7 & 1.6 & 7.5 & 9.1 & 6.2  \\
\hline
$\sigma (S_1 P_1)$ & 7.7 & 9.1 & 2.2 & 0.2 & 0.01  \\
\hline
$\sigma (S_2 P_1)$ & 8.3 & 4.9 & 0.09 & 0.001 & 4.5  \\
\hline
$\sigma (S_3 P_1)$ & 0.4 & 1.4 & 3.8 & 1.4 & 0.2  \\
\hline
$\sigma (S_1 P_2)$ & 0.7 & 1.5 & 3.6 & 1.4 & 0.2  \\
\hline
$\sigma (S_2 P_2)$ & 0.7 & 0.8 & 0.1 & 0.01 & 0.0006  \\
\hline
$\sigma (S_3 P_2)$ & 0.03 & 0.2 & 6.0 & 8.1 & 0.4  \\
\hline
$\sigma ({\bar \nu}_e \nu_e S_1)$ & 238.3 & 225.5 & 95.7 & 30.3 & 7.7 \\
\hline
$\sigma ({\bar \nu}_e \nu_e S_2)$ & 19.4 & 14.2 & 0.1 & 0.06 & 0.04  \\
\hline
$\sigma ({\bar \nu}_e \nu_e S_3)$ & 6.9 & 16.5 & 85.9 & 79.5 & 30.0  \\
\hline
$\sigma(e^+ e^- S_1)$ & 26.3 & 24.9 & 10.6 & 3.3 & 0.8 \\
\hline
$\sigma(e^+ e^- S_2)$ & 2.1 & 1.6 & 0.01 & 0.007 & 0.005 \\
\hline
$\sigma(e^+ e^- S_3)$ & 0.7 & 1.8 & 9.7 & 9.1 & 3.4 \\
\hline
\hline
\end{tabular}
\end{center}
\end{table}

\vfil\eject
\setcounter{figure}{0}
\def\figurename{}{}%
\renewcommand\thefigure{Fig. 1a}
\begin{figure}[t]
\epsfxsize=12cm
\hspace*{2.cm}
\epsffile{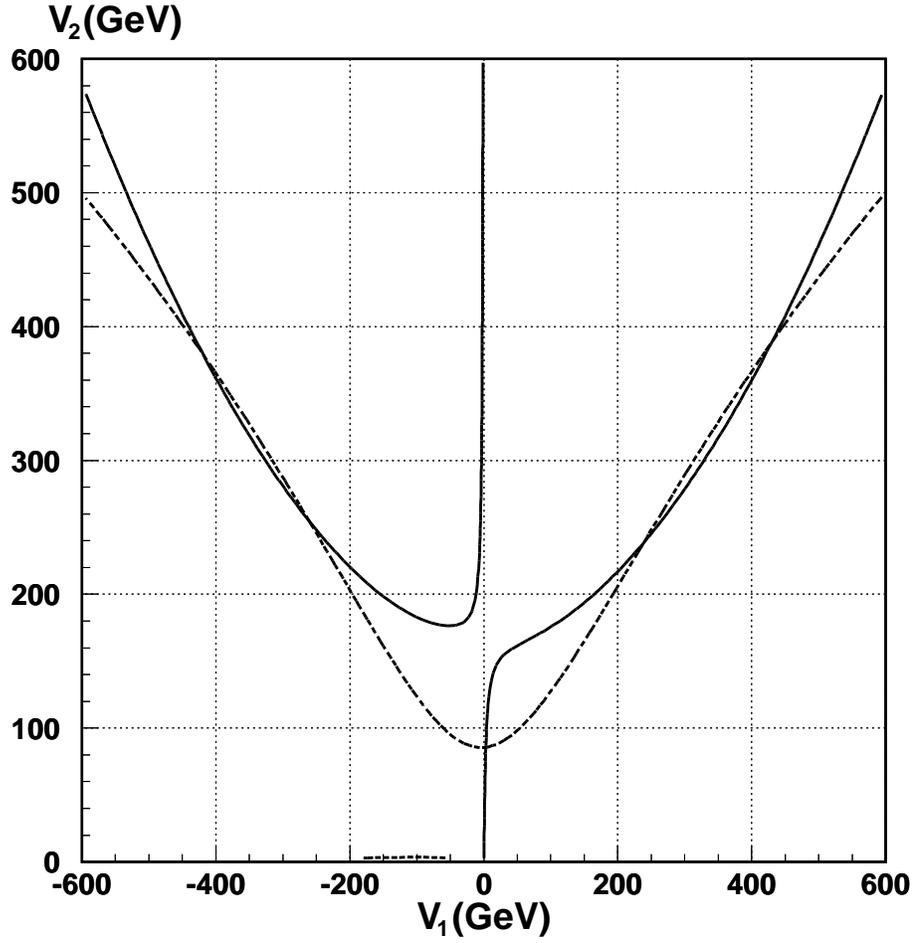}
\caption[plot]{Contours of $f_1 = 0$ (solid curve) and  $f_2 = 0$ (dashed curve)
in the $(v_1, v_2)$-plane for $\tan \beta = 2$ and $A_{\lambda} = 504.5$ GeV.
Other parameters are fixed as $\lambda = 0.3$, $x(0) = 25$ GeV, and $\xi = 50$ GeV.
The critical temperature is set as 100 GeV. Note that the two contours intersect
at five points in the $(v_1, v_2)$-plane.}
\end{figure}
\renewcommand\thefigure{Fig. 1b}
\begin{figure}[t]
\epsfxsize=12cm
\hspace*{2.cm}
\epsffile{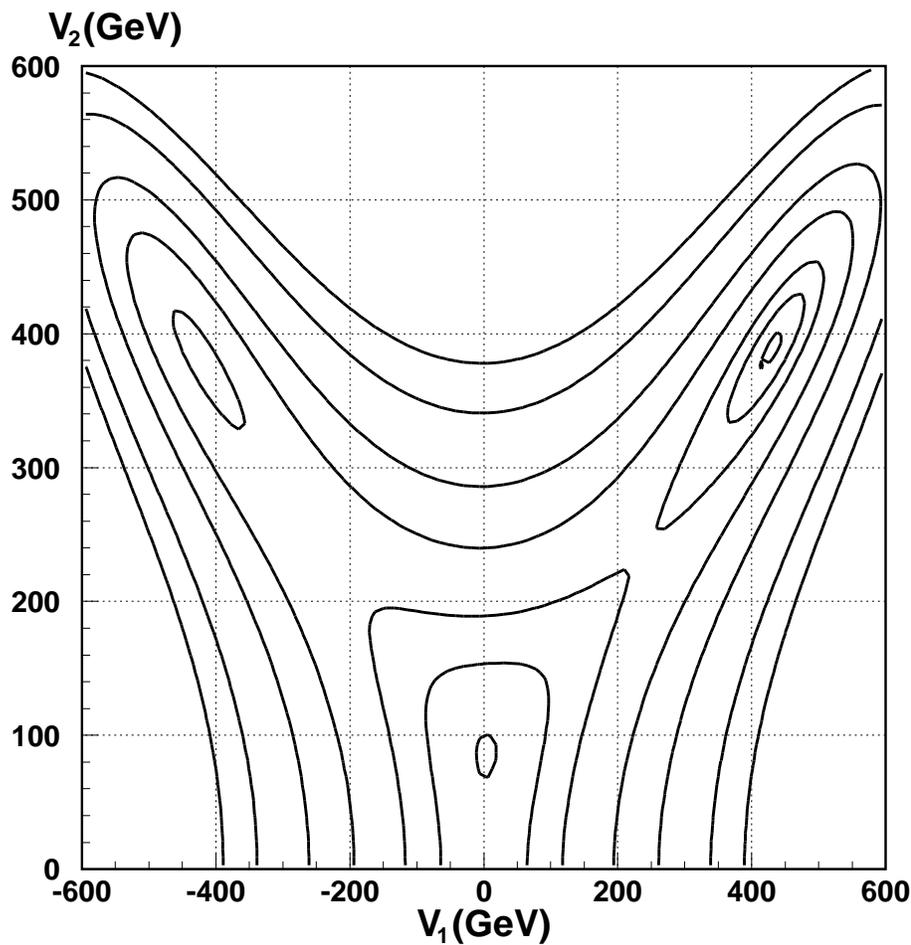}
\caption[plot]{Equipotential contours of $\langle V \rangle $ with the same parameter values as Fig. 1a.
There are five extrema; two of them are saddle points and the other three are minima.
Two of the three minima have the same value of $\langle V \rangle $,
thus define global minima or degenerate vacua.
The coordinates in unit of GeV of the degenerate vacua at point A and point B
in the $(v_1, v_2)$-plane the value of $x$ therefrom are respectively
(2.9, 83.1, 2.0) and (426.3, 385.0, 234.3). }
\end{figure}
\renewcommand\thefigure{Fig. 2a}
\begin{figure}[t]
\epsfxsize=12cm
\hspace*{2.cm}
\epsffile{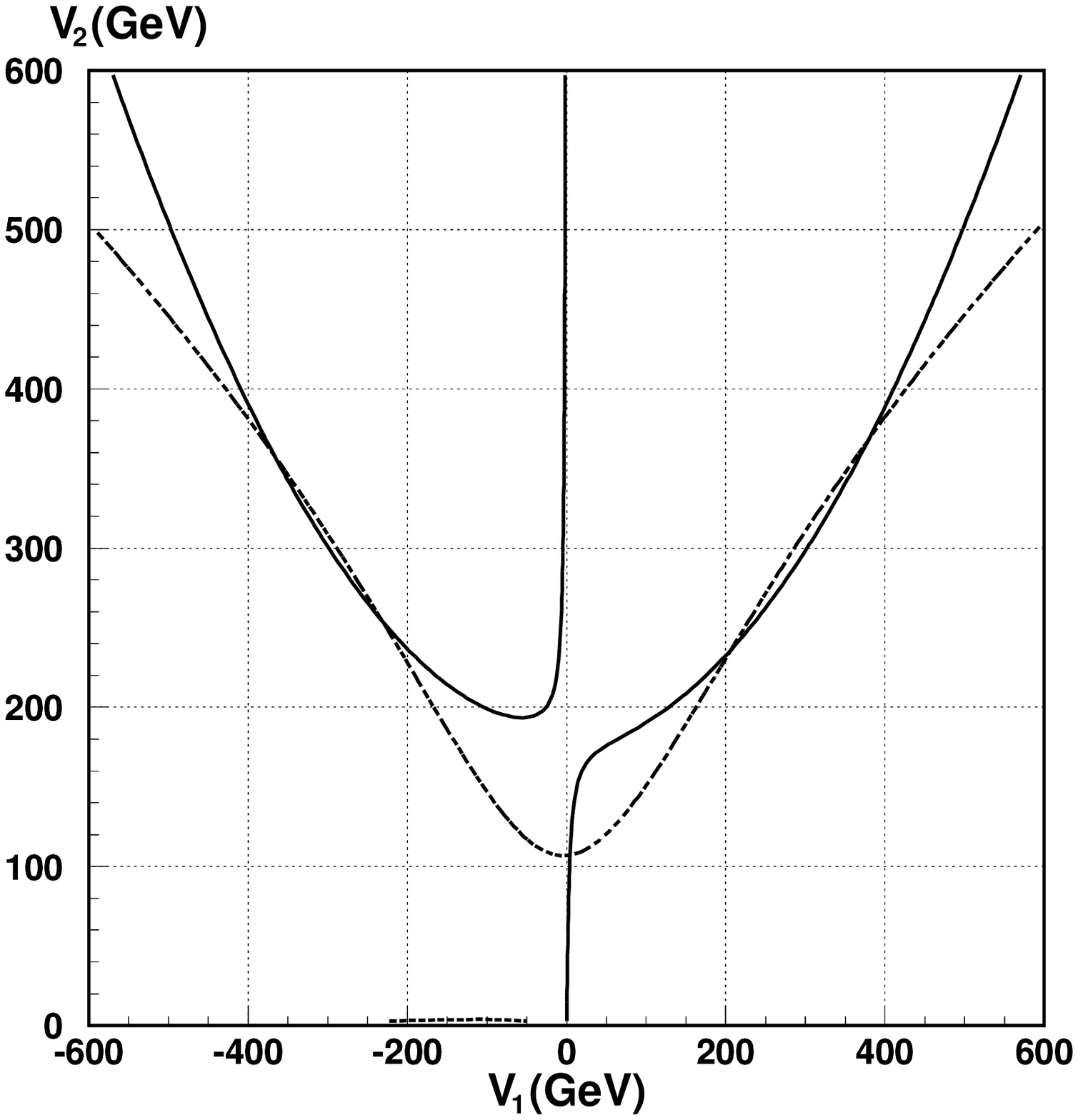}
\caption[plot]{The same as Fig. 1a, but with different values of $\tan\beta$ and
$A_{\lambda}$: $\tan \beta = 3$ and $A_{\lambda} = 433$ GeV.}
\end{figure}
\renewcommand\thefigure{Fig. 2b}
\begin{figure}[t]
\epsfxsize=12cm
\hspace*{2.cm}
\epsffile{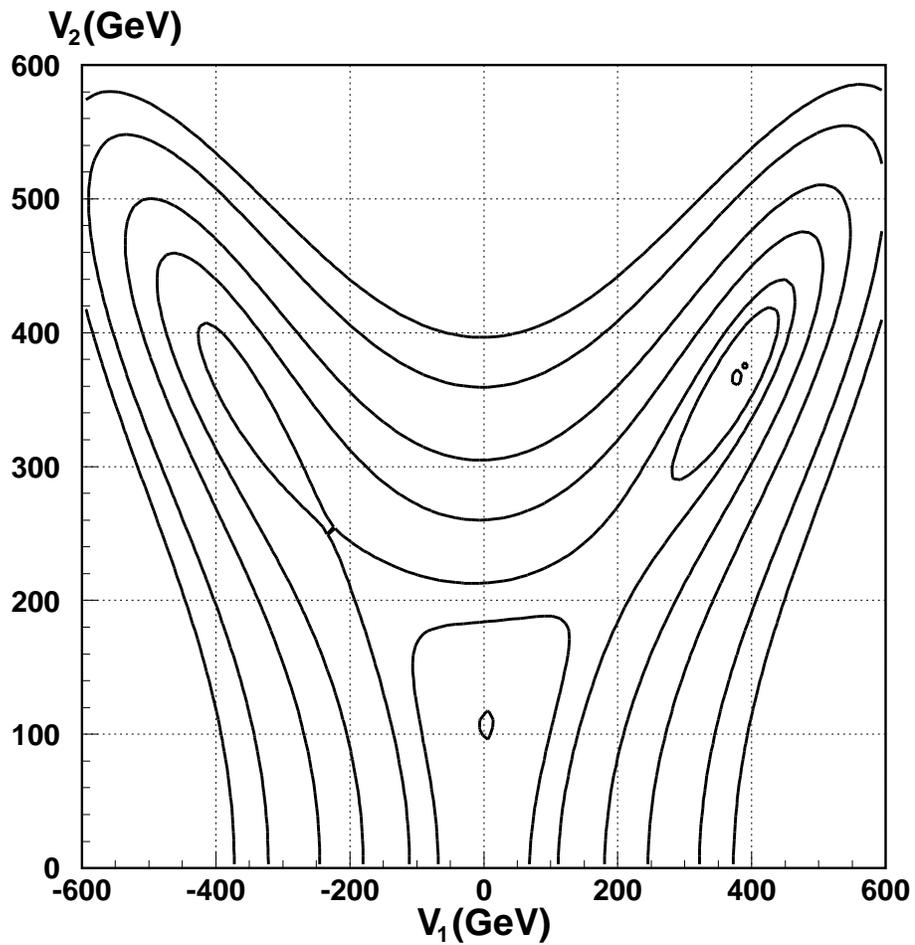}
\caption[plot]{The same as Fig. 1b, but with different values of $\tan\beta$
and $A_{\lambda}$: $\tan \beta = 3$ and $A_{\lambda} = 433$ GeV.
The coordinates in unit of GeV of the degenerate vacua in the $(v_1, v_2)$-plane
and the value of $x$ therefrom are respectively (4.2, 104.8, 3.5) and (375.7, 364.9, 238.7). }
\end{figure}
\renewcommand\thefigure{Fig. 3a}
\begin{figure}[t]
\epsfxsize=12cm
\hspace*{2.cm}
\epsffile{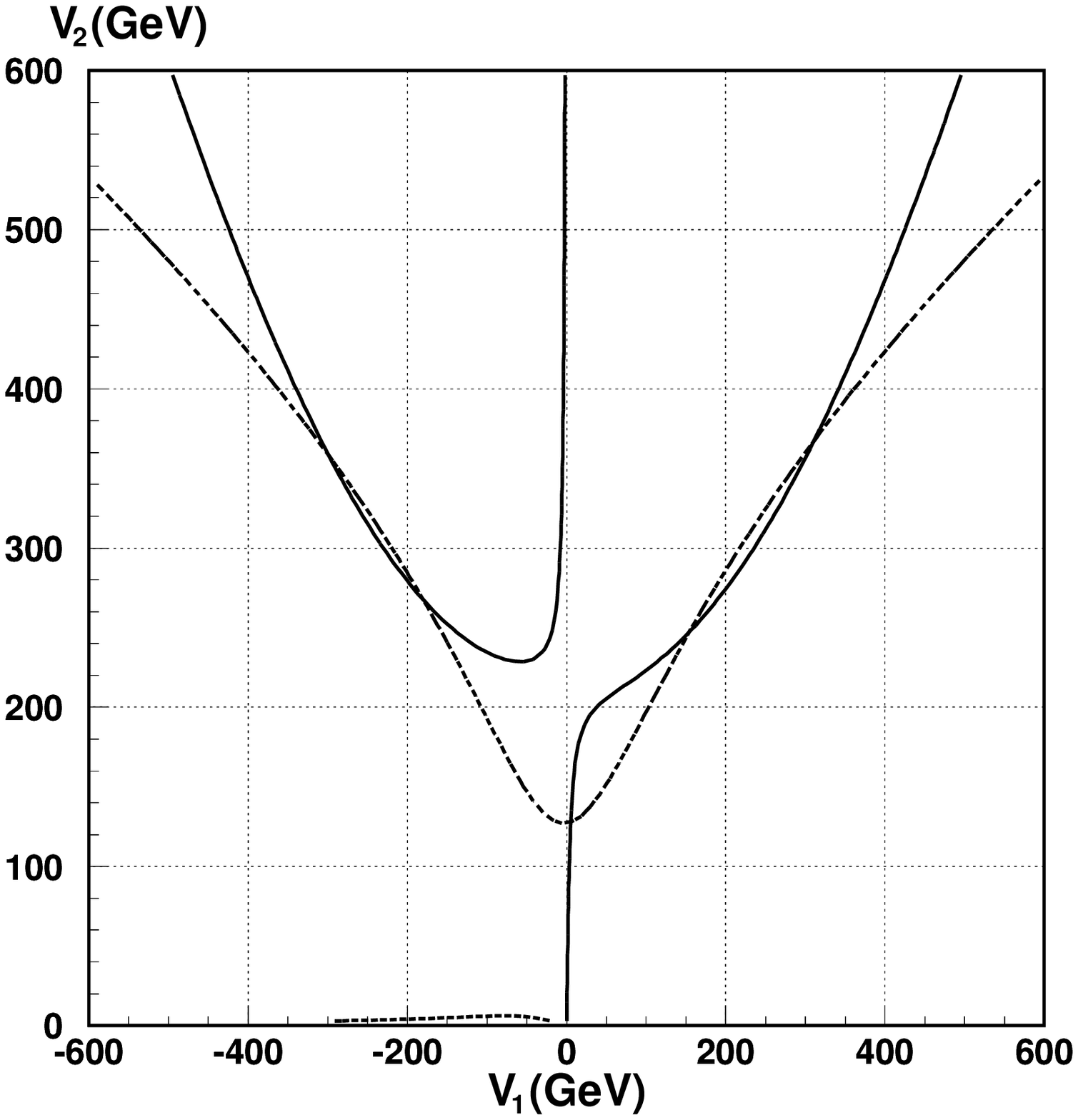}
\caption[plot]{The same as Fig. 1a, but with different values of $\tan\beta$
and $A_{\lambda}$: $\tan \beta = 10$ and $A_{\lambda} = 401$ GeV.}
\end{figure}
\renewcommand\thefigure{Fig. 3b}
\begin{figure}[t]
\epsfxsize=12cm
\hspace*{2.cm}
\epsffile{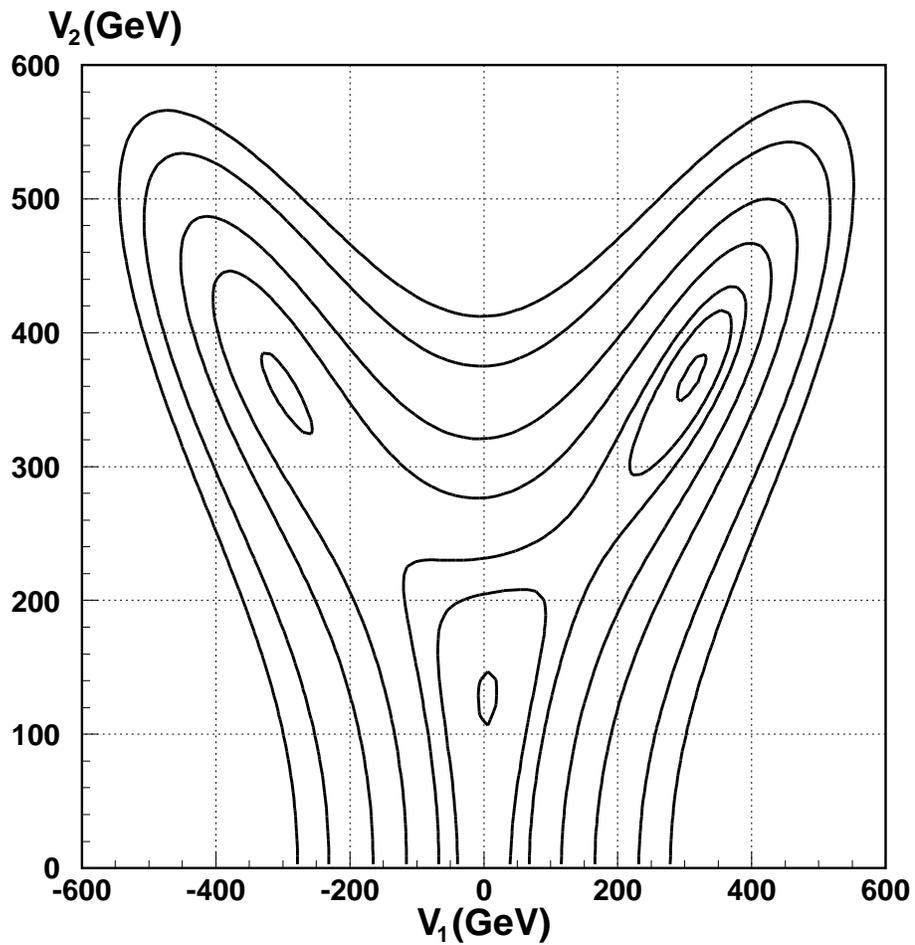}
\caption[plot]{The same as Fig. 1b, but with different values of $\tan\beta$ and
$A_{\lambda}$: $\tan \beta = 10$ and $A_{\lambda} = 401$ GeV.
The coordinates in unit of GeV of the degenerate vacua in the $(v_1, v_2)$-plane
and the value of $x$ therefrom are respectively (5.0, 125.8, 10.7) and (307.7, 364.7, 360.7). }
\end{figure}
\renewcommand\thefigure{Fig. 4a}
\begin{figure}[t]
\epsfxsize=12cm
\hspace*{2.cm}
\epsffile{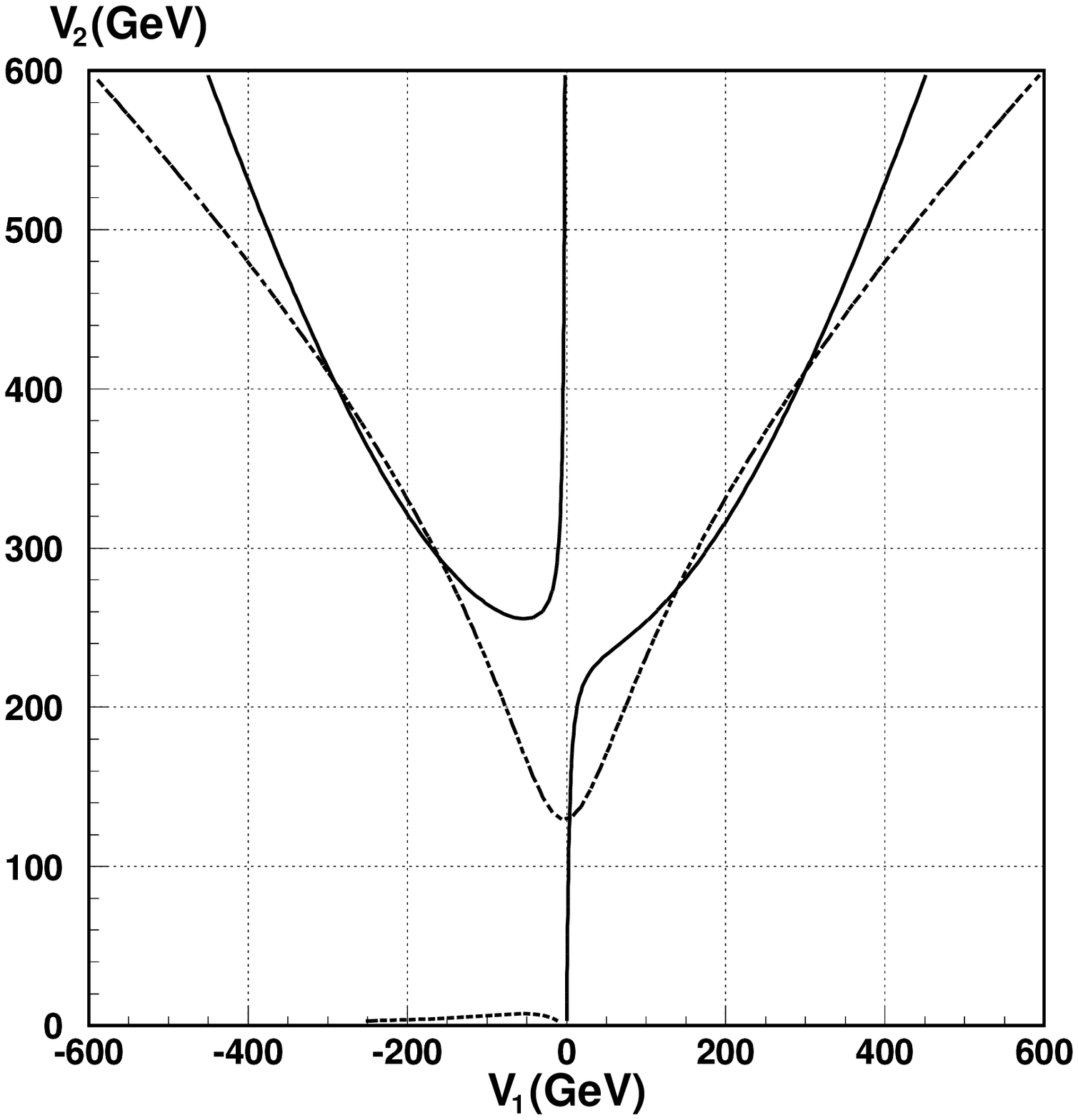}
\caption[plot]{The same as Fig. 1a, but with different values of $\tan\beta$
and $A_{\lambda}$: $\tan \beta = 20$ and $A_{\lambda} = 499$ GeV.}
\end{figure}
\renewcommand\thefigure{Fig. 4b}
\begin{figure}[t]
\epsfxsize=12cm
\hspace*{2.cm}
\epsffile{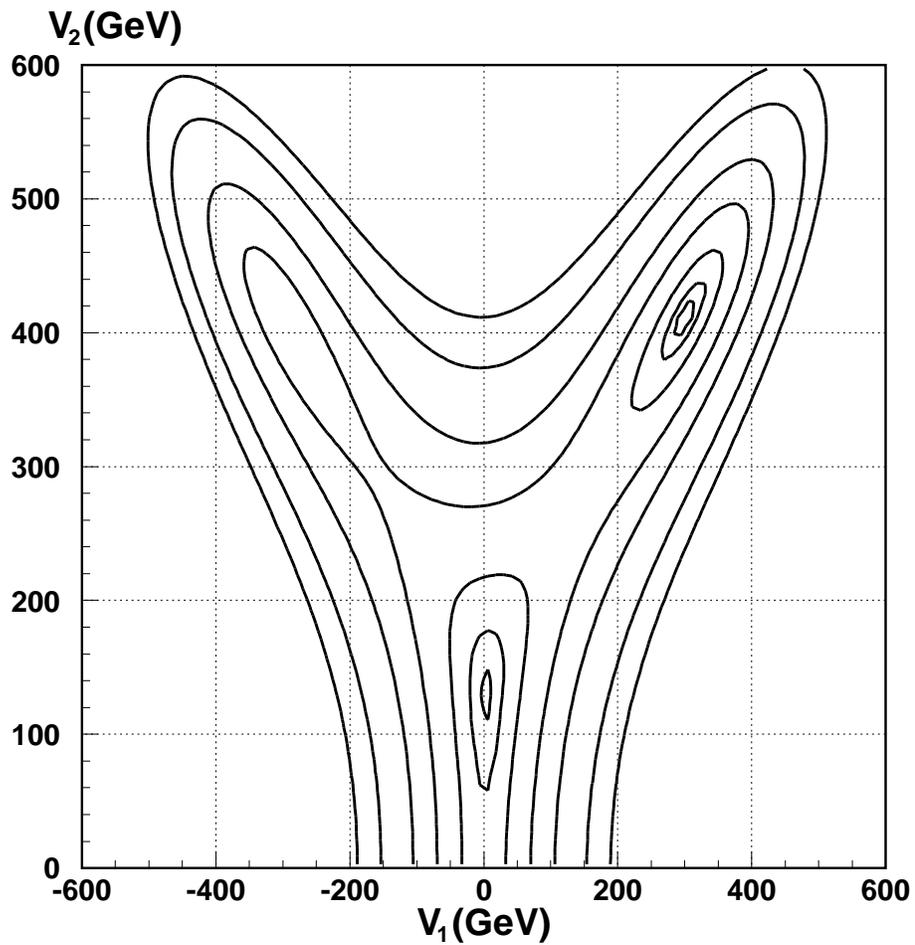}
\caption[plot]{The same as Fig. 1b, but with different values of $\tan\beta$
and $A_{\lambda}$: $\tan \beta = 20$ and $A_{\lambda} = 499$ GeV.
The coordinates in unit of GeV of the degenerate vacua in the $(v_1, v_2)$-plane
and the value of $x$ therefrom are respectively (3.6, 128.2, 14.6) and (298.0, 409.0, 526.6). }
\end{figure}
\renewcommand\thefigure{Fig. 5a}
\begin{figure}[t]
\epsfxsize=12cm
\hspace*{2.cm}
\epsffile{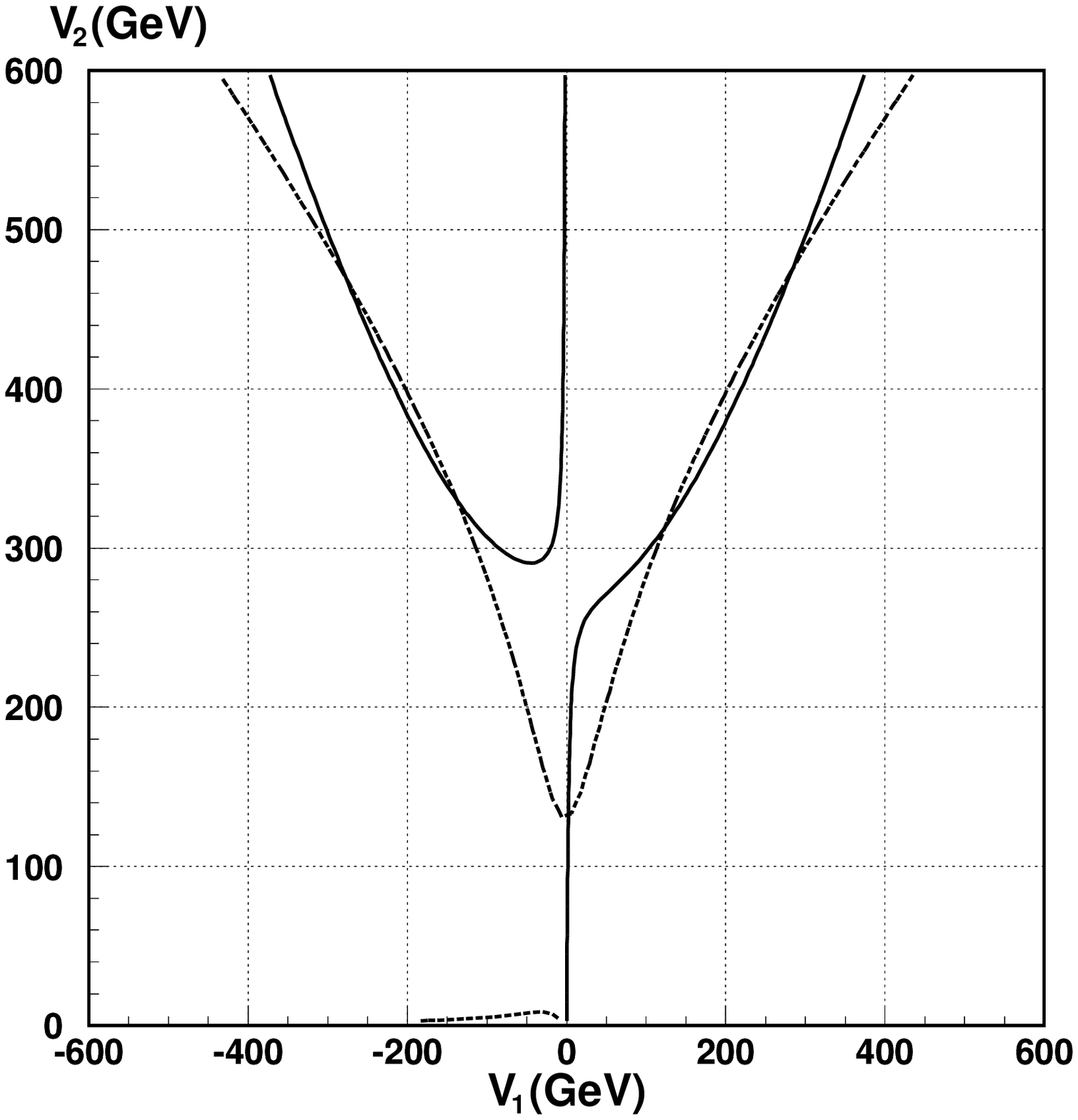}
\caption[plot]{The same as Fig. 1a, but with different values of $\tan\beta$
and $A_{\lambda}$: $\tan \beta = 40$ and $A_{\lambda} = 716.5$ GeV.}
\end{figure}
\renewcommand\thefigure{Fig. 5b}
\begin{figure}[t]
\epsfxsize=12cm
\hspace*{2.cm}
\epsffile{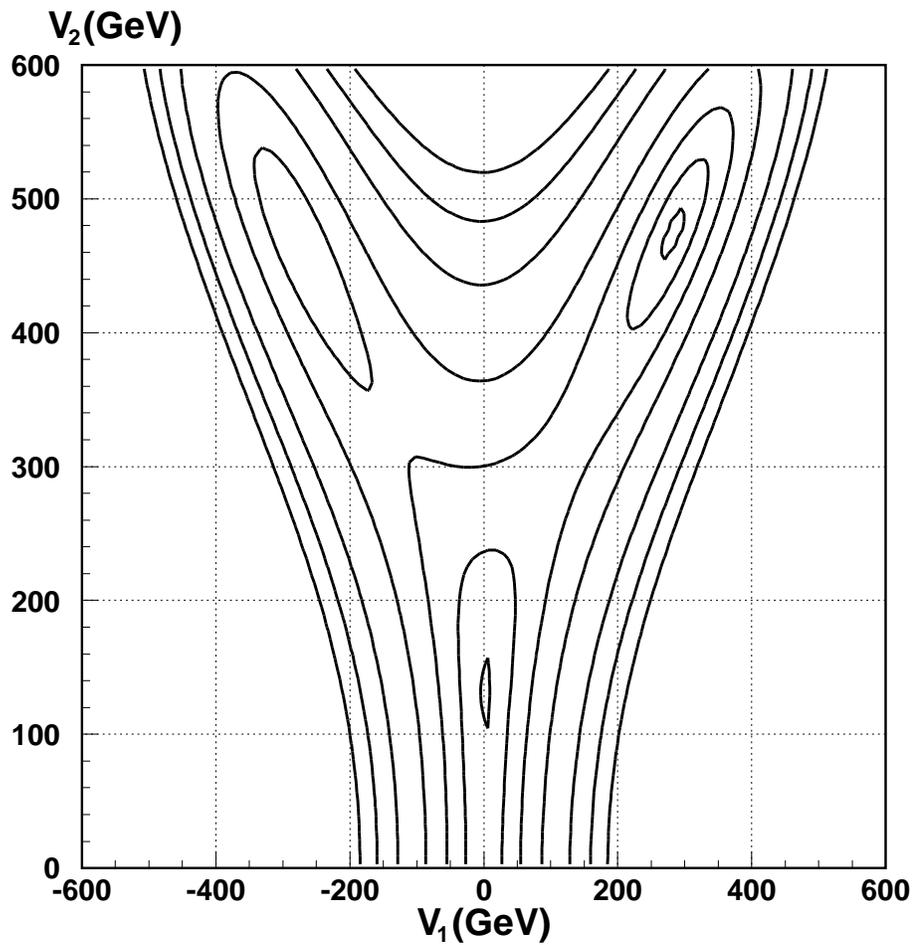}
\caption[plot]{The same as Fig. 1b, but with different values of $\tan\beta$
and $A_{\lambda}$: $\tan \beta = 40$ and $A_{\lambda} = 716.5$ GeV.
The coordinates in unit of GeV of the degenerate vacua in the $(v_1, v_2)$-plane
and the value of $x$ therefrom are respectively  (2.1, 129.0, 17.2) and (281.5, 472.7, 786.7). }
\end{figure}
\renewcommand\thefigure{Fig. 6}
\begin{figure}[t]
\epsfxsize=12cm
\hspace*{2.cm}
\epsffile{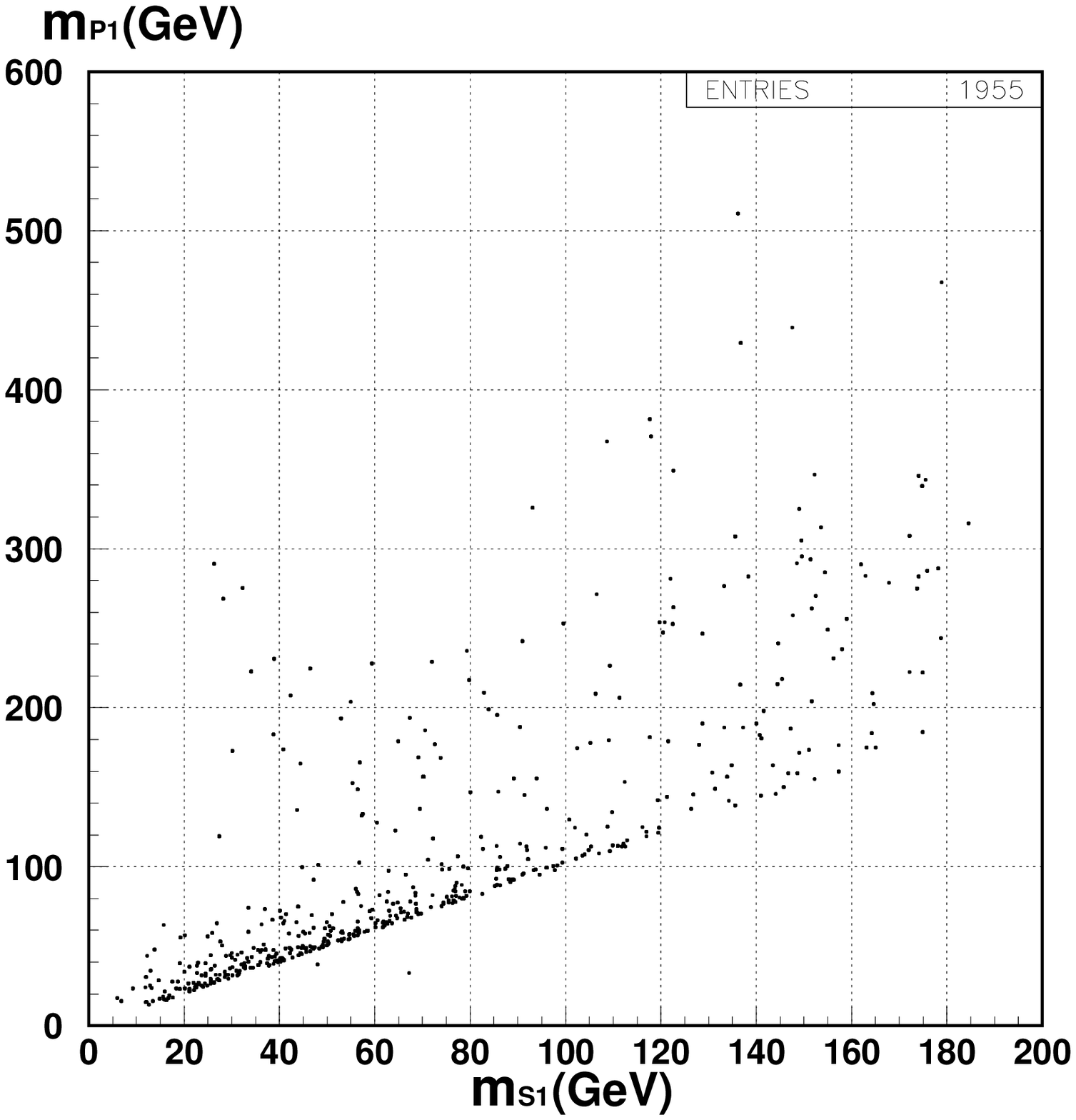}
\caption[plot]{Plot of the lightest scalar and pseudoscalar neutral Higgs boson masses
for the 1955 points in the parameter space of MNMSSM where the electroweak phase transition
is strongly first order. The points are selected in $2 \le \tan \beta \le 40$,
$0 < \lambda \le 0.7$, $0 < A_{\lambda} \le 1500$ GeV, and $0 < x(0), \xi \le 300$ GeV.}
\end{figure}
\renewcommand\thefigure{Fig. 7}
\begin{figure}[t]
\epsfxsize=12cm
\hspace*{2.cm}
\epsffile{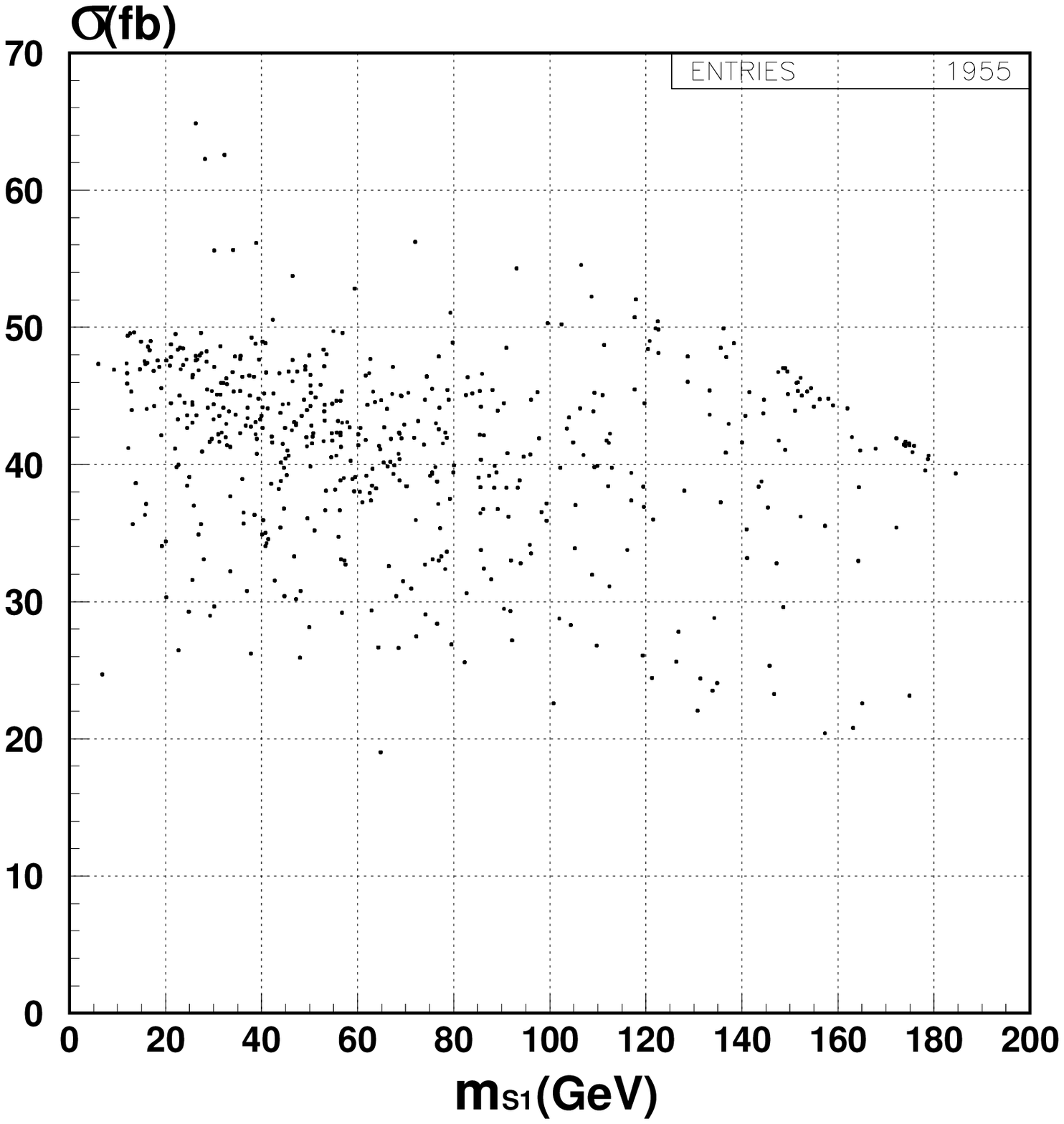}
\caption[plot]{Plot against the mass of the lightest scalar neutral Higgs boson of
the largest of the 15 cross sections for production of any neutral Higgs bosons
via any channel for the 1955 points in Fig. 6, in $e^+e^-$ collisions with $\sqrt{s} = 1000$ GeV. }
\end{figure}
\end{document}